\documentstyle[aps,epsfig,multicol,amssymb,amsfonts]{revtex}

\tighten
\renewcommand{\narrowtext} 
{\begin{multicols}{2}\global\columnwidth20.5pc} 
\renewcommand{\widetext}
{\end{multicols}\global\columnwidth42.5pc} 
\input psfig

\begin{document} 
\draft 
\title{Fluctuations of the inverse participation ratio at the Anderson
transition} 
\author{F.~Evers$^1$ and A.~D.~Mirlin$^{1,2,*}$ } 
\address{$^1$Institut f\"ur Theorie der Kondensierten Materie,
Universit\"at Karlsruhe, 76128 Karlsruhe, Germany}
\address{$^2$Institut
f\"ur Nanotechnologie, Forschungszentrum Karlsruhe, 76021 Karlsruhe,
Germany}
\date{\today}
\maketitle
\begin{abstract}
Statistics of the inverse participation ratio (IPR) at the critical
point of 
the localization transition is studied numerically for the power-law
random banded matrix model. It is shown that the IPR distribution
function is scale-invariant, with a power-law asymptotic
``tail''. This scale invariance implies that the
fractal dimensions $D_q$ are non-fluctuating
quantities, contrary to a recent claim in the literature. A
recently proposed relation between $D_2$ and the spectral
compressibility $\chi$ is violated in the regime of strong
multifractality, with $\chi\to 1$ in the limit $D_2\to 0$.
\end{abstract}

\narrowtext

Strong fluctuations of eigenfunctions represent one of the hallmarks
of the Anderson metal-insulator transition. These fluctuations can be
characterized by a set of inverse participation ratios (IPR)
\begin{equation}
P_q=\int d^dr\, |\psi({\bf r})|^{2q}\ .
\label{e1}
\end{equation}
In a pioneering work \cite{wegner80}, Wegner found from the
renormalization-group treatment of the $\sigma$-model in $2+\epsilon$
dimensions that the IPR show at criticality an anomalous scaling with
respect to the system size $L$,
\begin{equation}
\label{e2}
P_q\propto L^{-D_q(q-1)}\ .
\end{equation}
 Equation (\ref{e2}) should be contrasted
with the behavior of the IPR  in a good metal (where eigenfunctions
are ergodic), $P_q\propto L^{-d(q-1)}$, and, on the other hand, in the
insulator (localized  eigenfunctions),  $P_q\propto L^0$. 

The scaling (\ref{e2}) characterized by an infinite set of critical exponents
$D_q$ implies that the critical eigenfunction represents a
multifractal distribution \cite{castpel}. The notion of a multifractal
structure was 
first introduced by Mandelbrot \cite{mandelbrot74} and was later found
relevant in a variety of physical contexts, such as the
energy dissipating set in turbulence, strange attractors in chaotic
dynamical systems, and the growth probability distribution in
diffusion-limited aggregation; see \cite{paladin87} for a review. 
During the last decade, multifractality of critical eigenfunctions has
been a subject of intensive numerical studies \cite{janssen94}. 
Among all the multifractal dimensions, $D_2$
plays the most prominent role, since it determines the spatial
dispersion of the diffusion coefficient at the mobility edge
\cite{chalker88}. 

In fact, to make the statement (\ref{e2}) precise, one should specify
what exactly is meant by $P_q$ in its left-hand side. Indeed, the
IPR's  fluctuate from one eigenfunction 
(or one realization of disorder) to another. Should one take the average
$P_q$? Or, say, the most probable one? Will the results differ? More
generally, this poses the question of the form of the IPR distribution
function at criticality.

In a recent Letter \cite{parshin99}, Parshin and Shober addressed this
problem via numerical simulations for the 3D  tight-binding
model. Their main finding is that the fractal dimension $D_2$ is not a
well defined quantity, but rather shows universal 
fluctuations characterized by some distribution function ${\cal
P}(D_2)$ of a width of order unity. If true, this would
force one to reconsider virtually all aspects of the multifractality
phenomenon, such as the notion of the singularity spectrum
$f(\alpha)$, the form of the eigenfunction correlations and of the
density response at the mobility edge etc. In view of such a challenge
to the common lore, the issue requires to be
unambiguously clarified. 

We begin by reminding the reader of the existent analytical results
concerning the IPR fluctuations. 
While the direct analytical study of the Anderson transition in 3D is
not feasible because of the lack of a small parameter, statistics of
energy levels and eigenfunctions in a metallic mesoscopic sample
(dimensionless conductance $g\gg 1$) can be studied systematically in
the framework of the supersymmetry method; see \cite{m-review} for a
review. Within this approach, the IPR fluctuations were studied
recently \cite{fm95a,prigal98,m-review}. In particular, the 2D
geometry was considered, which, while not being a true Anderson
transition point, shows many features of criticality, in view of the
exponentially large value of the localization length. It was found
that the distribution function of the IPR $P_q$ normalized to its
average value $\langle P_q\rangle$ has a scale invariant form. In
particular, the relative variance of this distribution
(characterizing its relative width) reads
\begin{equation}
{{\rm var}(P_q)/ \langle P_q\rangle^2}={Cq^2(q-1)^2/ \beta^2
g^2}\ ,
\label{e3}
\end{equation}
where $C \sim 1$ is a numerical coefficient determined
by the sample shape (and the boundary conditions), and $\beta =1$ or 2
for the case of unbroken (resp. broken) time reversal symmetry. It is
assumed here that the index $q$ is not too large,
$q^2\ll\beta\pi g$. These findings motivated 
the  conjecture \cite{fm95a} that the IPR distribution
at criticality has in general a universal 
form, i.e. that the distribution function 
${\cal P}(P_q/P_q^{\rm typ})$ is independent of the size $L$ in the
limit $L\to\infty$. Here $P_q^{\rm typ}$ is a typical value of the
IPR, which can be defined e.g. as a median \cite{shapiro86}  of the
distribution ${\cal P}(P_q)$.  Normalization of $P_q$ by its average
value $\langle P_q\rangle$ (rather than by the typical
value $P_q^{\rm typ}$) would restrict generality of the statement; see
the discussion below. Practically speaking, the conjecture of
Ref.~\cite{fm95a} is that the distribution function of the IPR
logarithm, ${\cal P}(\ln P_q)$ simply shifts along the $x$-axis with
changing $L$. In contrast, the statement of Ref.~\cite{parshin99} is
that the width of this distribution function scales proportionally to
$\ln L$. 

While the above-mentioned analytical results for the 2D case are
clearly against the statement of \cite{parshin99}, their applicability
to a generic Anderson transition point may be questioned. Indeed, the
2D metal represents only an ``almost critical'' point, and the
consideration is 
restricted to the weak disorder limit $g\gg 1$ (weak coupling regime
in the field-theoretical language), while all the realistic
metal-insulator  transitions (conventional Anderson transition in 3D,
quantum Hall transition etc.) take place in the regime of strong
coupling. 

To explore the IPR fluctuations at criticality in the strong coupling
regime, we have performed numerical simulations of the power-law
random banded matrix (PRBM) ensemble. This model of the Anderson
critical point introduced in \cite{prbm} is defined as the ensemble of
random Hermitean $N\times N$ matrices $\hat H$ 
(real for $\beta=1$ or complex for $\beta=2$). 
The matrix elements $H_{ij}$ are independently distributed
Gaussian variables with zero
mean $\langle H_{ij}\rangle=0$ and the variance 
\begin{equation}
\label{e4}
\langle |H_{ij}|^2\rangle =a^2(|i-j|)\ ,
\end{equation}
where $a(r)$ is given by
\begin{equation}
\label{e5}
a^2(r)=\left[1+{1\over b^2}{\sin^2(\pi r/N)\over(\pi/N)^2}\right]^{-1}\ .
\end{equation}
Here $0<b<\infty$ is a parameter characterizing the ensemble, whose
significance will be discussed below. The crucial feature of the function
$a(r)$ is its $1/r$--decay for $r\gg b$. Indeed, for $r\ll N$
Eq.~(\ref{e5}) reduces to
\begin{equation}
a^2(r)=[1+(r/b)^2]^{-1}\ .
\label{e6}
\end{equation}
The formula (\ref{e5}) is just a periodic generalization of
(\ref{e6}), allowing to diminish finite-size effects (an analog of
periodic boundary conditions). 

In a straightforward interpretation, the model describes a 1D sample
with random long-range hopping, the hopping amplitude decaying as
$1/r$ with the length of the hop. Also, such an ensemble arises as an
effective description in a number of physical contexts.
Referring the reader to Refs.~\cite{prbm,m-review} for details (see
also \cite{kravtsov97,altshuler97,kravtsov99}), we
only give a brief summary of the main relevant analytical findings. The
PRBM model formulated above is critical at arbitrary value of $b$; it
shows all the key features of the Anderson critical point, including
multifractality of eigenfunctions and non-trivial spectral
compressibility (to be discussed below). Perhaps, the most appealing
property of the ensemble is the existence of the parameter $b$ which
labels the critical point: Eqs.~(\ref{e4}),
(\ref{e5}) define a whole family of critical theories parametrized
by $b$ \cite{bandcenter}. 
This is in full analogy with the family of the conventional
Anderson transition critical points parametrized by the spatial
dimensionality $2<d<\infty$. The limit $b\gg 1$ is 
analogous to $d=2+\epsilon$ with $\epsilon\ll 1$; it allows a
systematic analytical treatment (weak coupling expansion for the
$\sigma$-model). The opposite limit $b\ll 1$ corresponds to $d\gg 1$,
where the transition takes place in the strong disorder (strong
coupling) regime, and is also accessible to an analytical treatment
\cite{tobe} using the method of \cite{levitov90}.
This makes the PRBM ensemble a unique laboratory for
studying general features of the Anderson transition. 
Criticality of the PRBM ensemble was recently confirmed in numerical
simulations for $b=1$ \cite{varga99}. 
%Needless to say,
%changing $b$ in numerical simulations of the PRBM ensemble is much
%simpler than changing the spatial dimensionality $d$.

We have calculated the distribution function of the
IPR in the case $\beta=1$ for system sizes ranging
from $N=256$ to $N=4096$ and for
various values of $b$ by numerically diagonalizing the
Hamiltonian matrix defined in Eq. (\ref{e4})
using standard techniques. The statistical average is over a few
thousand matrices in the case of large system sizes up to
$10^5$ matrices at $N=256$. Specifically,
we have considered an average over wavefunctions having energies
in a small energy interval about the band center,
with a width of about $10\%$ of the band width. 

\begin{figure}
\includegraphics[width=0.8\columnwidth,clip]{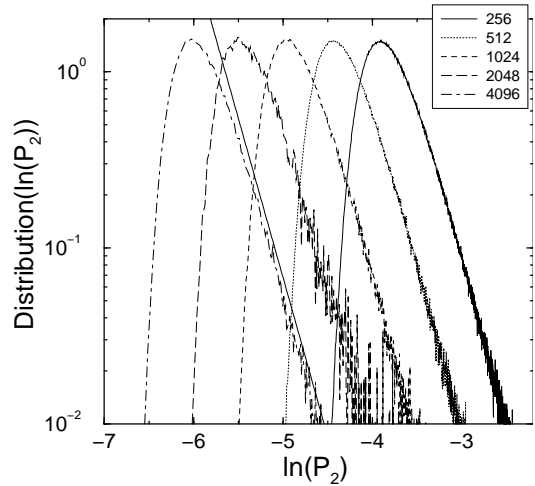}
%\vspace{3mm}
\caption{Distribution of $\ln P_2$ at $b=1$ for the system size
$N=256$, 512, 1024, 2048, and 4096. The straight line corresponds
to the power-law asymptotics with the index $x_2\approx 4.16$.}
\label{fig1} 
\end{figure}
 
Fig.~\ref{fig1} displays our result for the distribution of the IPR
logarithm,  ${\cal P}(\ln P_2)$.
It is clearly seen that the distribution function
does not change its shape or width with increasing $N$. 
After shifting the curves along the $x$-axis, they all lie on
top of each other, forming a scale-invariant IPR distribution. Of
course, the far tail of this universal distribution becomes
increasingly better developed with increasing $N$. 
From the shift of
the distribution ${\cal P}(\ln P_2)$ with $N$ we find the fractal
dimension $D_2=0.75\pm0.05$. Analogous results are obtained for other
values of $b$ and $q$ and will be published
elsewhere \cite{tobe}.

We conclude therefore that the distribution of IPR (normalized to 
its typical value) is indeed scale-invariant, in agreement with the
conjecture of Ref.~\cite{fm95a} and in disagreement with
Ref.~\cite{parshin99}. A natural question that can be asked 
is why the authors of \cite{parshin99} failed to find this
universality? We speculate that, possibly, the system sizes $L$ 
used in their numerical simulations were too small for observing the
universal form of ${\cal P}[\ln (P_2/P_2^{\rm typ})]$ in the limit
$L\to \infty$ \cite{levitov99}.  

\begin{figure}
\includegraphics[width=0.8\columnwidth]{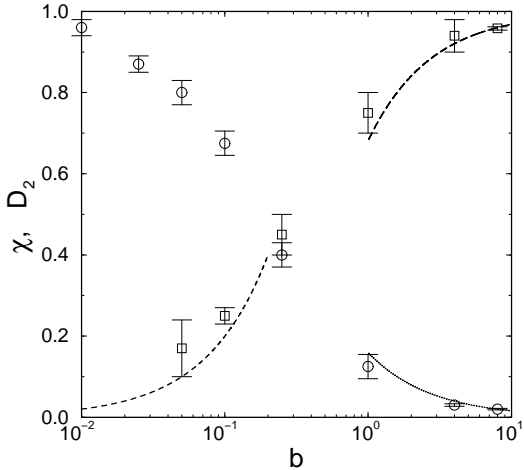}
%\vspace{3mm}
\caption{The fractal dimension $D_2$ (squares) and the spectral
compressibility $\chi$ (circles) as a function of the parameter $b$ of
the PRBM model. The corresponding $b\gg 1$ and $b\ll 1$ analytical
asymptotics are shown.}  
\label{fig2} 
\end{figure}

The value of the fractal dimension $D_2$ found from the scaling of the
shift of the distribution with $N$ is shown in Fig.~\ref{fig2} as a
function of the parameter $b$ of the PRBM model.
The numerical results agree very well with the analytical
asymptotics in the limits of large $b$, $\eta\equiv 1-D_2 =1/\pi b$ 
\cite{prbm,m-review} and small $b$, $D_2=2b$ \cite{tobe}.
We have also calculated the spectral compressibility $\chi$
characterizing fluctuations of the number $n$ of energy levels in a
sufficiently large energy window $\delta E$, 
%(which should be much larger than
%the level spacing but much smaller than the energy band width),
${\rm var}(n)=\chi \langle n\rangle$.  The results are also shown in
Fig.~\ref{fig2}, and are in perfect agreement with the
large-$b$ asymptotics, $\chi=1/2\pi b$ \cite{prbm,m-review}, as well. 
A non-trivial value of the spectral compressibility $0<\chi<1$
(intermediate between $\chi=0$ in a metal and $\chi=1$ in an
insulator) has been understood to be an intrinsic feature of the
critical point of the Anderson transition \cite{chi}. 

In a remarkable recent work \cite{chalker96}, Chalker, Lerner and
Smith employed Dyson's idea of Brownian motion through the
ensemble of Hamiltonians to link the spectral statistics with
wavefunction correlations. On this basis, it was argued in
Ref.~\cite{ckl} that the following exact relation between 
$\chi$ and $D_2$ holds:
\begin{equation}
\label{e7}
\chi=(d-D_2)/2d\ .
\end{equation}
According to (\ref{e7}), the spectral compressibility
should tend  to $1/2$ in the limit $D_2\to 0$ (very sparse
multifractal), and not to the Poisson value $\chi=1$.
However, the numerical data of Fig.~\ref{fig2}
show that, while being 
an excellent approximation at large $b$ (we remind that for our
system $d=1$), the relation (\ref{e7})
gets increasingly stronger violated with decreasing $b$. In
particular, in the limit $b\to 0$ (when $D_2\to 0$)
the spectral compressibility tends to the Poisson limit $\chi\to 1$. 
The same conclusion was reached analytically in \cite{m-review} for
the PRBM model with broken time reversal invariance. 
Similar violation of (\ref{e7}) is indicated by numerical data for the 
tight-binding model in dimensions $d>4$ \cite{zhar-private}. It would
be interesting to see why the derivation of (\ref{e7}) in \cite{ckl} 
fails at small $b$. 

Let us now comment on the necessity to distinguish between the average
value $\langle P_q\rangle$ and the typical value $P_q^{\rm typ}$. This
is related to the question of the asymptotic behavior of the
distribution ${\cal P}(P_q)$ at anomalously large $P_q$. It was found
in the 2D case \cite{m-review} that the distribution has a power-law
tail  ${\cal P}(P_q)\propto P_q^{-1-x_q}$ with $x_q=2\beta\pi g/q^2$
(as before, $g\gg 1$ and $q^2\ll\beta\pi g$ assumed). 
We believe that the power-law asymptotics with some $x_q>0$ is a
generic feature of the Anderson transition point. This is confirmed by
our numerical simulations, as 
illustrated in Fig.~\ref{fig1}. For not too large $q$
the index $x_q$ is sufficiently large ($x_q>1$), so that  
there is no essential
difference between  $\langle P_q\rangle$ and $P_q^{\rm typ}$. However,
with increasing $q$ the value of $x_q$
decreases. Once it drops below unity, the average $\langle P_q\rangle$
starts to 
be determined by the upper cut-off of the power-law ``tail'',
determined by the system size. As a result, for $x_q<1$ the average
shows a scaling  $\langle P_q\rangle\propto L^{-\tilde{D}_q(q-1)}$
with an exponent $\tilde{D}_q$ different from $D_q$ as defined from
the scaling of $P_q^{\rm typ}$ (see above). In this situation
the average value  
$\langle P_q\rangle$ is not representative and is determined by rare
realizations of disorder. Therefore, the condition $x_q=1$ corresponds
to the point $\alpha_-$ of the singularity spectrum with
$f(\alpha_-)=0$.  
If one performs the ensemble averaging in the regime $x_q<1$, 
one finds $\tilde{D}_q$ as the fractal
exponent and (after the Legendre transform) the 
function $f(\alpha)$ continuing beyond the point $\alpha_-$ into the
region $f(\alpha)<0$ \cite{m-review}. With this definition, the
fractal exponent $\tilde{D}_q\to 0$ as $q\to\infty$. On the other
hand, the fractal exponent $D_q$ defined above from the scaling of the
typical value $P_q^{\rm typ}$ (or, equivalently, of the whole
distribution function) corresponds to the spectrum $f(\alpha)$
terminating at $\alpha=\alpha_-$ and saturates
$D_q\to\alpha_-$ in the limit $q\to\infty$.

In the region $x_q>1$ (corresponding to $f(\alpha)>0$) the two
definitions of the fractal exponents are identical,
$D_q=\tilde{D}_q$. This is in particular valid at $q=2$ for the
Anderson transition in 3D and for the Quantum Hall transition. 

\begin{figure}
\includegraphics[width=0.8\columnwidth,clip]{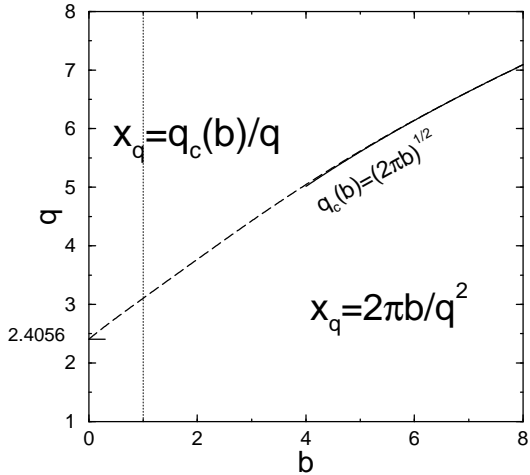}
%\vspace{3mm}
\caption{``Phase diagram'' of the multifractal spectrum. The phase
boundary $q_c(b)$ separates the regions of $x_q>1$ (below) and $x_q<1$
(above). The dotted line separates the asymptotic regimes $b\ll 1$ and
$b\gg 1$, for which the analytical results have been obtained
\protect\cite{tobe}. The dashed line is a schematic illustration of
the crossover between the two asymptotics.}  
\label{fig3} 
\end{figure}

As has been mentioned above, the two limits $b\gg 1$ and $b\ll 1$ can
be studied analytically. Let us announce the corresponding results for
the IPR statistics; details will be published elsewhere
\cite{tobe}. As shown in Fig.~\ref{fig3}, the ``phase boundary''
$q_c(b)$ separating the regimes of $x_q>1$  ($D_q=\tilde{D}_q$) and
$x_q<1$ ($D_q>\tilde{D}_q$) has the asymptotics $q_c=(2\pi b)^{1/2}$
($b\gg 1$) and $q_c=2.4056$ ($b\ll 1$). Notice that this implies
$D_2=\tilde{D}_2$ for all $b$. The corresponding power-law tail
exponent $x_2$ is equal to $\pi b/2$ at $b\gg 1$ and to $3/2$ at $b\ll
1$. The values of $x_q$ at $q<q_c$ (for $b\gg 1$) as well at $q>q_c$
are given in Fig.~\ref{fig3}. 

Finally, it is worth mentioning that the meaning of universality of
the IPR distribution at the critical point is 
the same as for the conductance distribution or for the level
statistics. Specifically, the IPR distribution does depend on the
system geometry (i.e., on the shape and on the boundary
conditions). However, for a given geometry it is independent of the
system size and of microscopic details of the model, and is an
attribute of the relevant critical theory.

In conclusion, we have studied the IPR statistics in the family of the
PRBM models of the Anderson transition. Our main findings are as
follows: (i) The distribution function of the IPR (normalized to its
typical value $P_q^{\rm typ}$) is scale-invariant, as was conjectured
in \cite{fm95a}. (ii) The scaling of $P_q^{\rm typ}$ with the system size
defines the fractal exponent $D_q$, which is a non-fluctuating
quantity, in contrast to \cite{parshin99}. (iii) The universal
distribution ${\cal P}(z\equiv P_q/P_q^{\rm typ})$ has a power-law
tail $\propto z^{-1-x_q}$. At sufficiently large $q$ 
one finds $x_q<1$, and the average
value $\langle P_q\rangle$ becomes non-representative and scales with
a different exponent $\tilde{D}_q\ne D_q$. (iv) The relation
(\ref{e7}) between the spectral compressibility and the fractal
dimension $D_2$ argued to be exact in Ref.~\cite{ckl} is violated in
the strong-multifractality regime. In particular, $\chi\to 1$ in the
limit of a very sparse multifractal ($D_2\to 0$).

Discussions with V.E.~Kravtsov, L.S.~Levitov, D.G.~Polyakov, I.~Varga
and I.Kh.~Zharekeshev are
gratefully acknowledged. This work was supported by the SFB 195 der
Deutschen Forschungsgemeinschaft.

\end{multicols}


\begin{references}
\bibitem{bandcenter} We concentrate on the band center $E=0$. Changing
the energy $E$ at fixed $b$ also allows to go through the
one-parametric family of the critical points; this way is, however,
numerically inconvenient.
\end{references}
\end{document}